\documentclass[fleqn,usenatbib,useAMS]{mnras}

\usepackage{graphicx}	%
\usepackage{amsmath}	%
\usepackage{amssymb}	%
\usepackage{times}

\clearpage{}%
\def\aj{AJ}%
\def\apj{ApJ}%
\def\aap{A\&A}%
\def\mnras{MNRAS}%
\def\nat{Nature}%
\def\physrep{Phys.~Rep.}%
\clearpage{}%

\title[Filaments From Weak Lensing]
	{The Weak Lensing Masses of Filaments between Luminous Red Galaxies}
\author[Epps \& Hudson]
	{Seth~D. Epps$^1$ \& 
	Michael~J.~Hudson$^{1,2}$ \thanks{Email: \href{mailto:mike.hudson@uwaterloo.ca}{mike.hudson@uwaterloo.ca}} \\
	$^1$Department of Physics \& Astronomy, 
	University of Waterloo, Waterloo, ON, N2L 3G1, Canada. \\
	$^2$Perimeter Institute for Theoretical Physics, 31 Caroline St. N., Waterloo, ON, N2L 2Y5, Canada.
	}
	
\newcommand{\Msun}{M_{\odot}}

\begin{document}
\bibliographystyle{mn2e}

\label{firstpage}

\maketitle

\nocite{*}
\begin{abstract}
In the standard model of non-linear structure formation, a cosmic web of dark-matter dominated filaments connects dark matter halos. In this paper, we stack the weak lensing signal of an ensemble of filaments between groups and clusters of galaxies. Specifically, we detect the weak lensing signal, using CFHTLenS galaxy ellipticities, from stacked filaments between SDSS-III/BOSS luminous red galaxies (LRGs). As a control, we compare the physical LRG pairs with projected LRG pairs that are more widely separated  in redshift space. We detect the excess filament mass density in the projected pairs at the $5\sigma$ level, finding a mass of $(1.6 \pm 0.3) \times 10^{13} M_{\odot}$ for a stacked filament region 7.1 $h^{-1}$ Mpc long and 2.5 $h^{-1}$ Mpc wide. This filament signal is compared with a model based on the three-point galaxy-galaxy-convergence correlation function, as developed in \cite{2014arXiv1402.3302C}, yielding reasonable agreement.
\end{abstract}

\newpage
\section{Introduction}
A key prediction of the cold dark matter (CDM) model is that a network of low-density filaments connects dark matter halos. Measuring the signal from these structures is therefore a key part of understanding the large-scale structure in the universe. The most prominent of these diffuse filaments are expected to thread the most massive dark matter halos in the universe, where galaxy clusters will form. The existence of this filamentary structure is widely accepted, however there is limited direct observational evidence of these dark-matter dominated filaments. One of the best ways to probe the structure of dark matter is by weak gravitational lensing, where the distortion of background galaxies can be used to map out the foreground distribution of mass density.

Several authors have reported the detection of a dark matter filament connecting individual massive clusters using weak lensing. \cite{2012Natur.487..202D} found a dark matter filament connecting two massive ($\sim 10^{14} \Msun$) clusters, Abell 222 and Abell 223. 
More recently, \cite{2015arXiv1503.06373H} claimed the detection of a filament between the massive galaxy clusters CL0015.9+1609 and RX J0018.3+1618.
These individual filament detections rely on somewhat arbitrary parametric filament model that is difficult to interpret. \cite{2005MNRAS.359..272C} studied filaments between clusters in N-body simulations and found that, between clusters of galaxies separated by $\lesssim 10 h^{-1}$ Mpc, $\sim 90$\% are separated by filaments which have a typical cylindrical radius $\sim 2 h^{-1}$ Mpc. However, these filaments are not always straight, which complicated their identification even for massive filaments.

The weak lensing signal-to-noise of a single filament between a single pair of galaxy groups is expected to be much less than unity. The approach of this paper will be to stack many thousands of filaments between pairs of Luminous Red Galaxies (LRGs). LRGs inhabit halos of masses of a few times $10^{13} \Msun$ and so can be used as a proxy for galaxy groups \citep{ManSelCoo06}. When stacking filaments, the signal is best understood as the ensemble average of shear (or projected surface mass density) around halo pairs.  One way to model the stacked filament is through higher order perturbation theory, \emph{i.e.}, the three-point correlation function or bispectrum. The three point galaxy-galaxy-shear correlation function from weak lensing has been studied by a number of authors \citep{2003MNRAS.340..580T, 2005A&A...432..783S, 2008A&A...479..655S, 2014arXiv1402.3302C}

Recently, \cite{2013MNRAS.430.2476S} used CFHTLens data and measured three-point statistics of galaxy number density and convergence. From this extracted the excess surface mass density around stacked lens galaxy pairs, both early-type and late-type. They found an excess surface mass density around early-type lens galaxy pairs, with the excess around late-type pairs being consistent with zero. This analysis used of photometric redshifts to identify pairs of galaxies.  As will be discussed in \S\ref{sec:sdss} below, the disadvantage is that the relatively large error in photometric redshifts ($\sim 0.05$ or $\sim 150h^{-1}\mathrm{Mpc}$) will scatter physically connected pairs of galaxies away, and scatter seemingly independent pairs together, and so complicates the interpretation of the results.

Clampitt and collaborators \citep{2014arXiv1402.3302C, ClaMiyJai16} have investigated the stacked weak lensing signal between SDSS LRGs at various separations, based on SDSS spectroscopy and imaging. 
\cite{2014arXiv1402.3302C} presented two filament models, one based on the three-point correlation function, and the other a string of Navarro-Frenk-White \citep[hereafter NFW]{1997ApJ...490..493N} halos. The published version of the same paper \citep{ClaMiyJai16} instead compared the data with stacked filaments from N-body simulations, finding reasonable agreement. The latter paper reported a detection at the $4.5\sigma$ level, although no mass was quoted for the filament.

In this work, we describe techniques needed to measure the stacked filament between groups and clusters of galaxies, and apply these to LRG pairs. We also the attempt to model the filament using the three-point correlation function. 
In Section \ref{sec:data}, we discuss the data: CFHTLenS for galaxy source ellipticities and photometric redshifts, and the Baryon Oscillation Spectroscopic Survey \citep[hereafter BOSS]{2013AJ....145...10D} for spectroscopic redshifts of LRGs, a proxy for groups and cluster centres.  The LRG-pair stacking procedure is outlined in Section \ref{sec:measure}, and the results are presented in both shear and convergence maps. We also introduce the  technique of subtracting non-physical pairs in order to isolate the filament signal from the shear signal of the individual clusters. Finally, we provide an empirical measurement of the stacked filament surface mass density and total mass. In Section \ref{sec:3PCF}, we describe a  model for stacked filament in the context of the perturbation theory, starting from the three-point galaxy-galaxy-convergence correlation function. We compare this model with the data, and discuss possible improvements to the model.  
Section \ref{sec:conc} summarizes our results
Throughout this work we adopt a cosmology with the following parameters:  
$\Omega_{\mathrm{m}}=0.3$,
$\Omega_{\Lambda} = 0.7$,
$h\equiv H_0/(100\mathrm{km\,s}^{-1}\mathrm{Mpc}^{-1})=0.7$,
$n_s = 0.96$,
and $\sigma_8 = 0.8$. 

\section{Data}\label{sec:data}
In order to study the weak lensing signal of filaments one requires two sets of data: a catalogue of galaxy groups and cluster lens pairs, and a catalogue of background source galaxies with accurate ellipticity measurements.
\subsection{CFHTLenS background source galaxies}\label{sec:cfhtlens}
The CFHTLenS data were derived from the Wide component of the Canada-France-Hawaii Telescope Legacy Survey (CFHTLS), which was optimized for weak lensing measurements.  Observations were taken between March 2003 and November 2008 with the CFHT MegaPrime instrument which has roughly a $1^{\circ}\times 1^{\circ}$ field of view. The CFHTLS Wide data includes photometry in five optical passbands ($u^{*}, g^{\prime}, r^{\prime}, i^{\prime}, z^{\prime}$) and covers $\sim 154$ square degrees in four patches on the sky (W1-W4), 3 of which have substantial overlap with BOSS/SDSS-III as discussed below.  The  deepest band ($i^{\prime}$) data yields 17 resolved galaxies per square arcminute \citep{2013MNRAS.433.2545E}.

Galaxy ellipticity measurements were obtained with the `\emph{lensfit}' algorithm \cite{2013MNRAS.429.2858M}, modelled with bulge and disk components ultimately giving the two ellipticity parameters, $e_{\mathrm{1}}$ and $e_{\mathrm{2}}$ by Bayesian marginalization over galaxy size, centroid and bulge fraction. A corresponding \emph{lensfit} weight was assigned to each galaxy given the variance of the ellipticity likelihood surface defined in equation 8 of \cite{2013MNRAS.429.2858M}. After weighting, the effective source density is 11 galaxies per square arcminute \citep{2012MNRAS.427..146H}.

Photometric redshifts (photo-$z$s) were estimated using the Bayesian Photometric Redshift (BPZ) code outlined in \cite{2000ApJ...536..571B}, making use of the five-band photometry available from CFHTLS \citep{2012MNRAS.421.2355H}, yielding a mean photometric redshift of 0.75, much deeper than the lens sample of $\sim 0.4$. The photo-$z$s are limited to the range $0.2 < z_{\mathrm{phot}} < 1.3$, with a scatter of $\sigma_{z} \sim 0.04(1+z)$ and a catastrophic outlier rate of $\lesssim 4\%$ \citep{2012MNRAS.427..146H}. For a detailed description of the methods used to estimate the photo-$z$s, see \cite{2012MNRAS.421.2355H}.
\subsection{Lenses: SDSS LRG Pairs}\label{sec:sdss}
In N-body simulations, filaments connect the high density nodes where galaxy groups and clusters will be forming.%
To identify pairs of galaxy groups and clusters that are connected by a filaments, one requires an accurate estimate of their location in redshift space. Unfortunately, the uncertainty associated with photometric redshifts will scatter true physical pairs away from each other and scatter false projected pairs to the same redshift. For example, if there are two physically-associated galaxies are scattered by a photometric redshift of $\Delta z_{\mathrm{phot}} = 0.05$ (the typical photo-$z$ uncertainty in CFHTLenS), the corresponding scatter in their line-of-sight separation would be $\sim 150h^{-1}\mathrm{Mpc}$. This is much larger than the physical line-of-sight separation of order $\sim 10h^{-1}\mathrm{Mpc}$.  To mitigate this issue, physical pairs should be identified with spectroscopic redshifts, which have orders of magnitude better redshift accuracy ($\sigma_{z_{\mathrm{spec}}} \sim 10^{-4}$ or $\sigma_{v} \sim 30$ km/s).

BOSS has obtained spectroscopic redshifts for a large sample of LRGs, an excellent proxy for the centres of galaxy groups and clusters.
In this study both the BOSS CMASS and LOWZ sample galaxies were selected using the color-magnitude cuts from \cite{2013AJ....145...10D}. The majority of the overlap on the sky between the BOSS and CFHTLenS surveys is in the W1, W3 and W4 patches \citep[see]{2015ApJ...806....1M}, and giving $\sim 24,000$ LRGs in total.

A catalogue of LRG pairs was constructed by selecting pairs that were separated in redshift by $\Delta z_{\mathrm{spec}} < 0.002$ (corresponding to $\sim 5 h^{-1}$ Mpc comoving if in the Hubble flow), and separated in projection (i.e.\ on the sky) by $6 h^{-1}\mathrm{Mpc}\leq R_{\mathrm{sep}} < 10h^{-1}\mathrm{Mpc}$. This gave a sample of $\sim 23,000$ pairs of LRGs, with a mean physical separation of $\langle R_{\mathrm{sep}} \rangle \sim 8.23 h^{-1}\mathrm{Mpc}$, a mean redshift $\langle z \rangle \sim 0.42$, and a mean stellar mass of $\langle \log_{10}M_{\star}/M_{\odot}\rangle \sim 11.3$. According to \cite{2015MNRAS.447..298H}, these LRGs are expected to lie in halos of total mass  $\langle \log_{10}M/M_{\odot}\rangle = 13.04\pm0.07$, corresponding to galaxy groups.

\section{Measurement of Filament Signal}\label{sec:measure}

In \S\ref{sec:lensing}, we outline the technical details of stacking the shear signal from the lens-source system and describe our method for isolating the filament signal between the LRGs. In \S\ref{sec:emp_results} we present results for stacked LRG pairs.
\subsection{Lensing Shear Signal}\label{sec:lensing}
Unlike galaxy-galaxy lensing, where one is interested in the circularly averaged \emph{tangential} shear around individual galaxy centers, measuring the shear signal around pairs of LRGs is more complicated. The main complication arises because signal is not spherically symmetric, producing a shear signal that is not purely tangential. When stacking the lens-pair-source system,  it is necessary to keep track of both components of the source ellipticity, $e_1$ and $e_2$. In addition, one must account for the random orientations of LRG pairs, and their variable separation length. In \S\ref{sec:coords} below we develop a standardized coordinate system that allows for the stacking of arbitrary orientations and length, and in \S\ref{sec:stack} the actual stacking procedure is outlined.
\subsubsection{Standardized Coordinates}\label{sec:coords}
\begin{figure}
  \begin{center}
    \centerline{\includegraphics[width=1.2\linewidth]{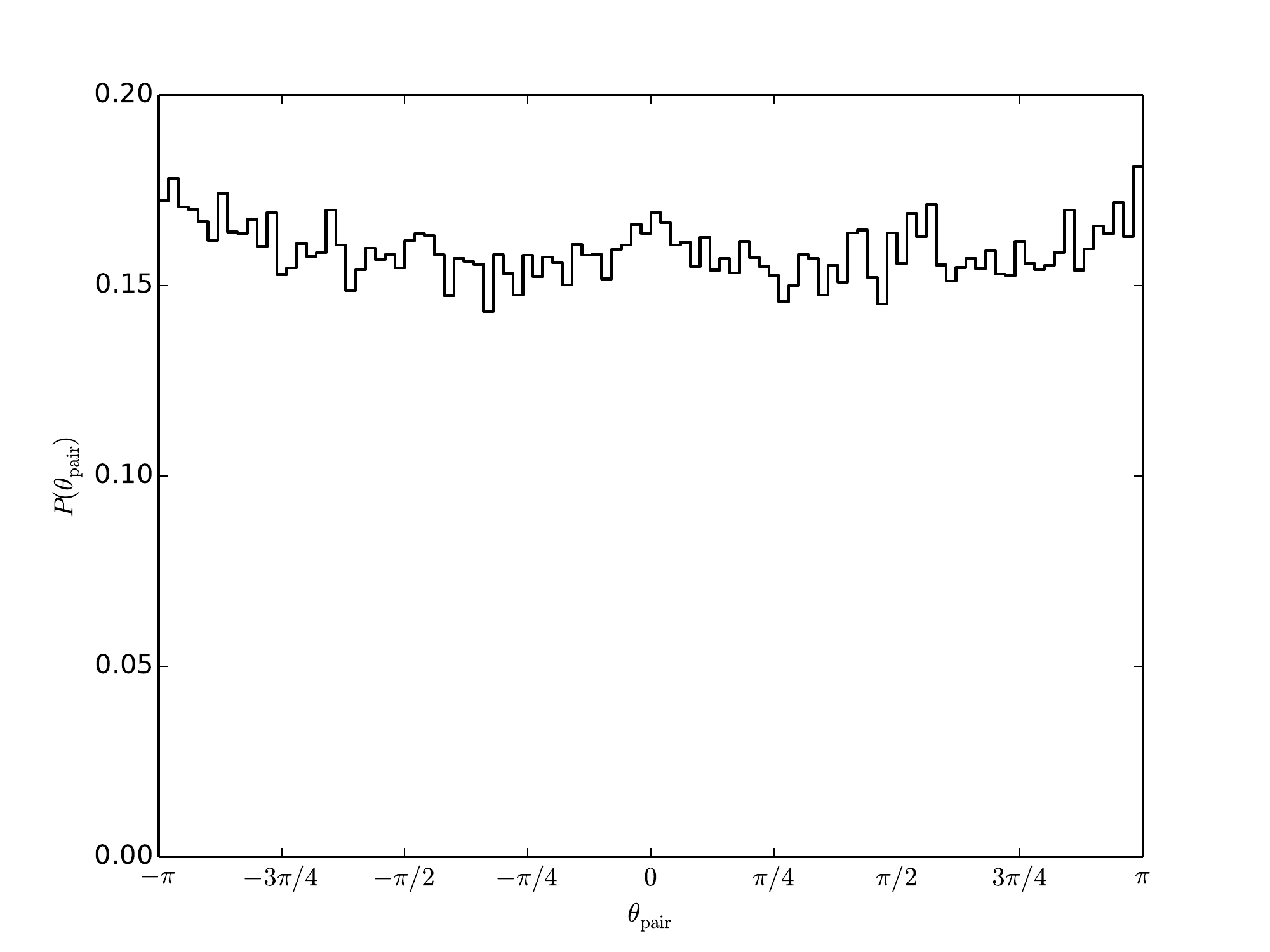}}
  \end{center}
  \caption[Distribution of LRG pair orientations]{The distribution of LRG pair angles, $\theta$, as measured in the tangent plane projection relative to the $X$-axis, see equation (\ref{eqn:theta_pair}). It is clear that the LRGs are distributed uniformly from $(-\pi,\pi)$} \label{fig:orient_dist}
\end{figure} 

In galaxy-galaxy lensing, one bins source galaxies in radial annuli around the lens centre. Here, however, we wish to stack LRG pairs which have uniform random orientations relative to the background galaxies (see Figure \ref{fig:orient_dist}), and varying physical separations. To account for this, we define a standardized coordinate system, normalized by pair separation, $R_{\mathrm{sep}}$, and rotated such that the LRG pair coordinates will translate to $(x_{L},y_{L})=(-0.5,0)$ and $(x_{R},y_{R})=(0.5,0)$. The source galaxies' positions and ellipticities must also be translated into this coordinate system as follows.
\begin{enumerate}
	\item {First the galaxy's position is translated such that the central right ascension and declination, $(\alpha_{\mathrm{c}}, \delta_{\mathrm{c}})$, of the LRG pair is at the origin, then projected into the tangent plane of the central point,
	\begin{eqnarray}\label{eqn:tangent_plane}
		X^{\prime}_{\mathrm{g}} &=& - (\alpha_{\mathrm{g}} - \alpha_{\mathrm{c}} )\cos{\delta_{\mathrm{c}}} \nonumber \\
		Y^{\prime}_{\mathrm{g}} &=& \delta_{\mathrm{g}} - \delta_{\mathrm{c}}.
	\end{eqnarray}
	}
	\item { Next the coordinates are rotated such that the LRG pair lies along the $x$-axis. This is done using the rotation matrix,
		\begin{equation}\label{eqn:rot_mat}
		R = \begin{bmatrix}
			\cos{\theta} & \sin{\theta} \\
			-\sin{\theta}  & \cos{\theta}
		\end{bmatrix},
		\end{equation}
		where $\theta$ is the angle between the individual LRGs about the central point in the tangent plane,
		\begin{equation}\label{eqn:theta_pair}
			\theta=\tan^{-1}\left ( \frac{Y^{\prime}_{R} - Y^{\prime}_{L}}{X^{\prime}_{R} - X^{\prime}_{L}} \right ).
		\end{equation}
		The subscripts $L,R$ represent the ``left" and ``right" LRGs in the pair.
	}
	\item {Finally the coordinates are rescaled by the separation between the two LRGs in the tangent plane, 
	\begin{equation}
		s = \sqrt{(\alpha_{R} - \alpha_{L})^{2}\cos^{2}{\delta_{\mathrm{c}}} + (\delta_{R} - \delta_{L})^{2} }.
	\end{equation} 
	This is the angular separation that corresponds to a projected physical separation, $R_{\mathrm{sep}}$.
	}
\end{enumerate}
Putting it all together, the final position of a galaxy in this coordinate system will be,
	\begin{eqnarray}\label{eqn:new_coord}
		x_{\mathrm{g}} &=& \frac{1}{s}\left [ (\alpha_{\mathrm{c}} - \alpha_{\mathrm{g}} )\cos{\delta_{\mathrm{c}}}\cos{\theta} + (\delta_{\mathrm{g}} - \delta_{\mathrm{c}} )\sin{\theta}         \right ] \nonumber \\
		y_{\mathrm{g}} &=& \frac{1}{s}\left [ -(\alpha_{\mathrm{c}} - \alpha_{\mathrm{g}} )\cos{\delta_{\mathrm{c}}}\sin{\theta} + (\delta_{\mathrm{g}} - \delta_{\mathrm{c}} )\cos{\theta}         \right ] 
	\end{eqnarray}
With the source galaxies in the new coordinate system, their ellipticities also need transformed. The 2 components of ellipticity only need to be rotated. The rotation matrix is nearly the same as (\ref{eqn:rot_mat}), however the property that ellipticity is invariant under $180^{\circ}$ rotation requires that the angle just be doubled,
\begin{eqnarray}\label{eqn:ellip}
		e^{\prime}_1 &=& e_1\cos{2\theta} + e_2\sin{2\theta}  \nonumber \\
		e^{\prime}_2 &=& -e_1\sin{2\theta} +  e_2\cos{2\theta}
\end{eqnarray}
\subsubsection{Stacking}\label{sec:stack}
The signal from an individual filament is expected to be very weak because the filament density is much lower than that of a galaxy or cluster of galaxies, so it is necessary to stack LRG pairs, \emph{i.e.} to take an ensemble average. To stack the source ellipticities around a pair of LRGs (from here on referred to as the `lens'), a two dimensional grid is prepared based on the $x-y$ coordinate system developed in \S\ref{sec:coords}. For each lens, at all $(x,y)$ cells of the grid, the the shear components are computed by averaging the source galaxy  ellipticities ($e_1$ and $e_2$) according to their \emph{lensfit} weights, $w$, with an additional factor of $\Sigma^{-2}_{\mathrm{crit}}$ as in \cite{2015MNRAS.447..298H}. The additional factor of $\Sigma^{-2}_{\mathrm{crit}}$ is used to down-weight sources that are near the lens in redshift, for which the signal is expected to be very weak. The critical surface density, $\Sigma_{\mathrm{crit}}$, is given by
\begin{equation}
	\Sigma_{\mathrm{crit}}(z_{\ell}, z_{j}) = \frac{c^2}{4\pi G} \frac{D(z_{j})}{D(z_{\ell})D(z_{\ell},z_{j})}, 
	\end{equation}
where $D(z_{\ell})$ is the angular diameter distance to the lens, $D(z_{j})$ is the angular diameter distance to the source, and $D(z_{\ell}, z_{j})$ is the angular diameter distance between the lens and source. To summarise, the ellipticities are stacked to obtain estimates of the shear according to
\begin{eqnarray}\label{eqn:shears}
		\gamma_1(x,y) &=& \frac{\sum_{\ell}\sum_{j\in(x,y)} e^{\prime}_{1,j}w_{j} 
			\Sigma^{-2}_{\mathrm{crit};\ell,j}}{\sum_{\ell}\sum_{j\in(x,y)}w_{j}\Sigma^{-2}_{\mathrm{crit};\ell,j}} \nonumber \\ \nonumber \\
		\gamma_2(x,y) &=& \frac{\sum_{\ell}\sum_{j\in(x,y)} e^{\prime}_{2,j}w_{j} 
			\Sigma^{-2}_{\mathrm{crit};\ell,j}}{\sum_{\ell}\sum_{j\in(x,y)}w_{j}\Sigma^{-2}_{\mathrm{crit};\ell,j}} ,
	\end{eqnarray}
where the average is over all lenses, $\ell$, and background sources, $j$, that belong to cell $(x,y)$ after the coordinate transformation. An additive correction is applied to the $e_{2}$ component (before rotating) when computing the shears, according equation (19) of \cite{2012MNRAS.427..146H}, that accounts for a bias in CFHTLenS lensfit ellipticity measurement. %
Additionally, \cite{2013MNRAS.429.2858M} found that a multiplicative correction for noise bias needs to be applied \emph{after} the ellipticities are stacked, calculated from
	\begin{equation}
		1 + K = \frac{\sum_{\ell}\sum_{j} \left [1 + m(\nu_{SNR}, r_{\mathrm{gal}})_{j}\right] w_{j} 
			\Sigma^{-2}_{\mathrm{crit};\ell,j}}{\sum_{\ell}\sum_{j}w_{j}\Sigma^{-2}_{\mathrm{crit};\ell,j}} .
	\end{equation}
The resulting corrected shears are then 
	\begin{equation}\label{eqn:calibrated}
		\gamma^{\mathrm{cor}}_{1,2}(x,y) =  \frac{\gamma_{1,2}(x,y)}{1+K}.
	\end{equation}
\subsubsection{Convergence \& Surface Mass Density}\label{sec:conv_SD}
One problem with examining shear maps directly is that they are difficult to interpret. Unlike the case of galaxy-galaxy lensing, where one can interpret the stacked tangential shears in terms of the mean excess mass density, in the case studied here, there is no analogous interpretation of the individual shear components. One solution is to use the method of \cite{1993ApJ...404..441K} to convert the shear map into a convergence map, which is proportional to the surface mass density in the lens plane. 
From the definition of convergence, we easily convert it to the surface mass density
	\begin{equation}
		\Sigma = \kappa\overline{\Sigma}_{\mathrm{crit}},
	\end{equation}
where $\overline{\Sigma}_{\mathrm{crit}}$ is the ensemble average, calculated using %
	\begin{equation}\label{eqn:SigCrit_field}
		\overline{\Sigma}_{\mathrm{crit}} = \frac{\sum_{\ell}\sum_{j}\Sigma_{\mathrm{crit};\ell,j} \cdot
			\Sigma^{-2}_{\mathrm{crit};\ell,j} w_j }{\sum_{j} w_j\Sigma^{-2}_{\mathrm{crit};\ell,j}}.
	\end{equation}
The mean $\overline{\Sigma}_{\mathrm{crit}}$ was found to be $1640 M_{\odot}/\mathrm{pc}^2$ for our sample.

\subsubsection{Isolating The Filament Signal}\label{sec:isolate}
The goal of this paper is to study the filaments that link groups and clusters. Filaments themselves are difficult to define. For our purposes, we will define the filament as the excess mass present in a pair of LRGs, over and above that expected from the individual haloes of the LRGs themselves. Therefore the contribution from the two LRGs must be removed. 
 One requires a method that will remove any tangential shear produced by the LRG halos, leaving behind a signal only from the filament. \cite{ClaMiyJai16} introduced an elegant nulling method based on combining shear data at four different points, rotated with respect to the two LRGs in such a way as to null the spherically symmetric part of the signal.  The disadvantage of their scheme is that the resulting signal combines signal from several locations and so it is difficult to visualize and understand.  In this paper, we opt for a simpler approach: compare physical LRG pairs with ``non-physical'' (projected) LRG pairs.

A particular pair of LRGs are likely to be physically connected if their line-of-sight separation is small. In this paper, we have adopted $\Delta z = 0.002$, corresponding to a line-of-sight separation $\sim 6 h^{-1}$ Mpc, to define physical pairs.   By contrast, the same approach can be used to find LRG pairs that have such a large line-of-sight separation that the probability of being connected by a filament is negligible. Such pairs only appear to be pairs in projection, and we shall refer to them as ``non-physical'' pairs.   Non-physical pairs of LRGs are selected to have a line-of-sight separation between 100$h^{-1}$Mpc and 120$h^{-1}$Mpc corresponding to a separation in redshift of $0.033 \lesssim \Delta z \lesssim 0.04 $. For determining background sources, we assume that the lens redshift is the average of the pair. When the ellipticities of sources that are behind the non-physical pairs are stacked, there should only be contributions from the two LRGs. Therefore by subtracting the stacked map of the non-physical pairs from that of the physical pairs, the remaining signal should be due to the filament. With this method, the data can be compared to the model in terms of shears or in terms of convergence ($\kappa$). Since it is easier to interpret the convergence signal, the remainder of the paper will focus on the $\kappa$ maps. 

\subsection{Results}\label{sec:emp_results}
\begin{figure*}
  \centering
  \resizebox{2.0\columnwidth}{!}{\includegraphics{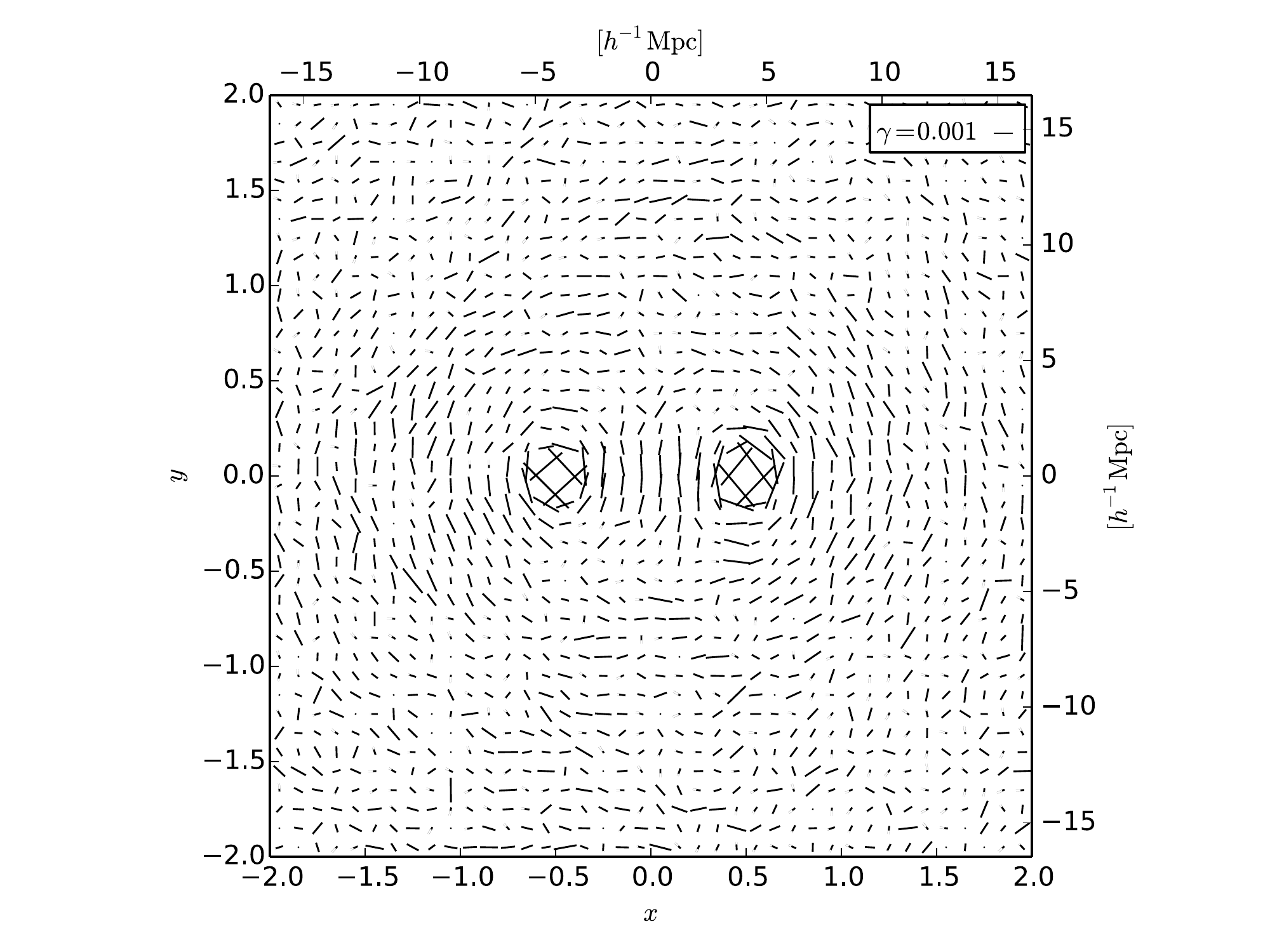}}
\caption[Shear map from stacking LRG pairs between $6h^{-1}\mathrm{Mpc} < R < 10h^{-1}\mathrm{Mpc}$ ]{The resulting shear map after stacking background galaxy ellipticities for LRG pairs with projected physical separations $6h^{-1}\mathrm{Mpc} < R < 10h^{-1}\mathrm{Mpc}$. The shears have been re-binned into a coarse grid for the purpose of illustration. The tangential nature of the shears around the LRGs is clearly visible. } \label{fig:shear_map}
\end{figure*}
\begin{figure*}
  \centering
  \resizebox{1.8\columnwidth}{!}{\includegraphics{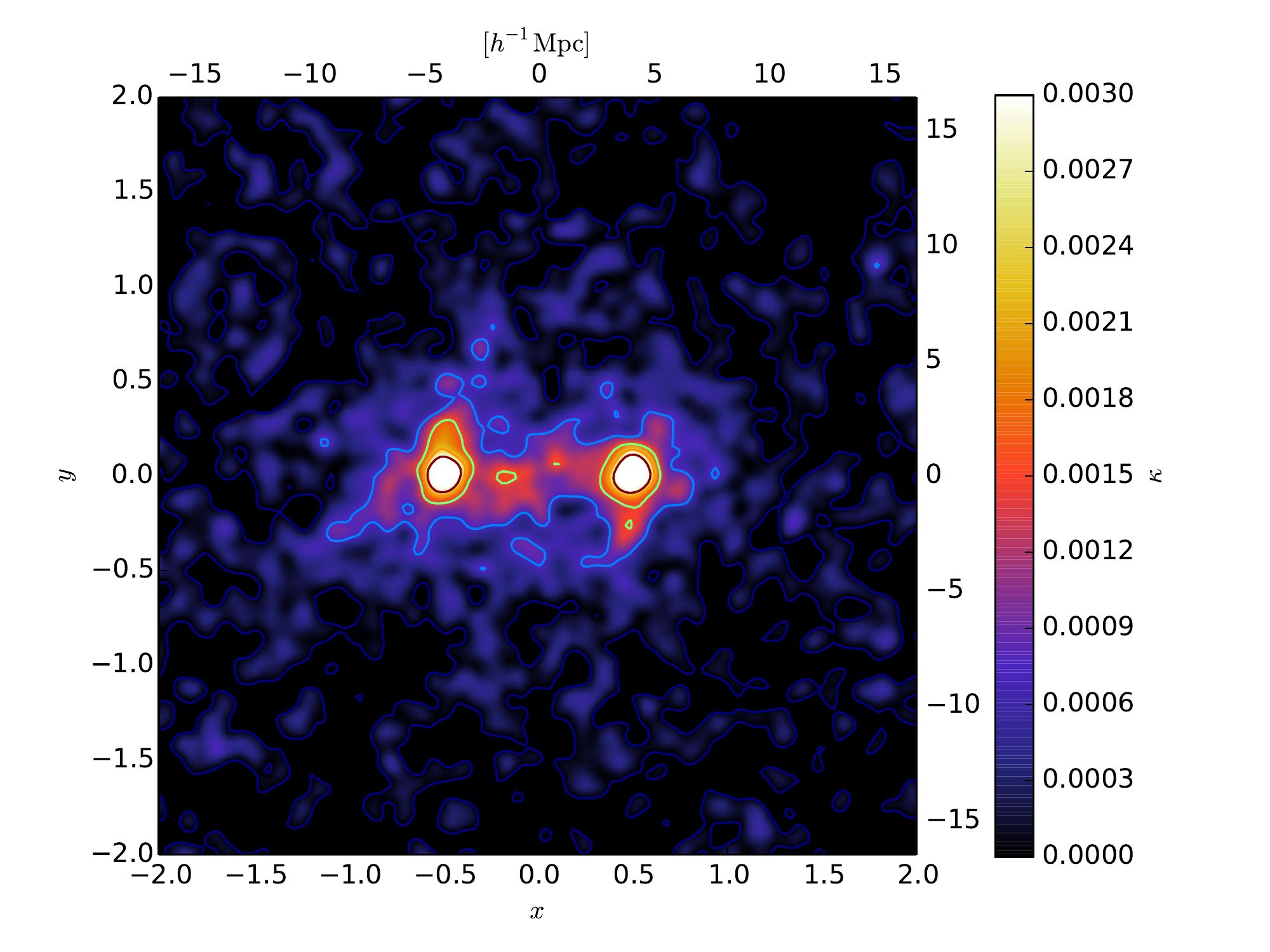}}\\
  \resizebox{1.8\columnwidth}{!}{\includegraphics{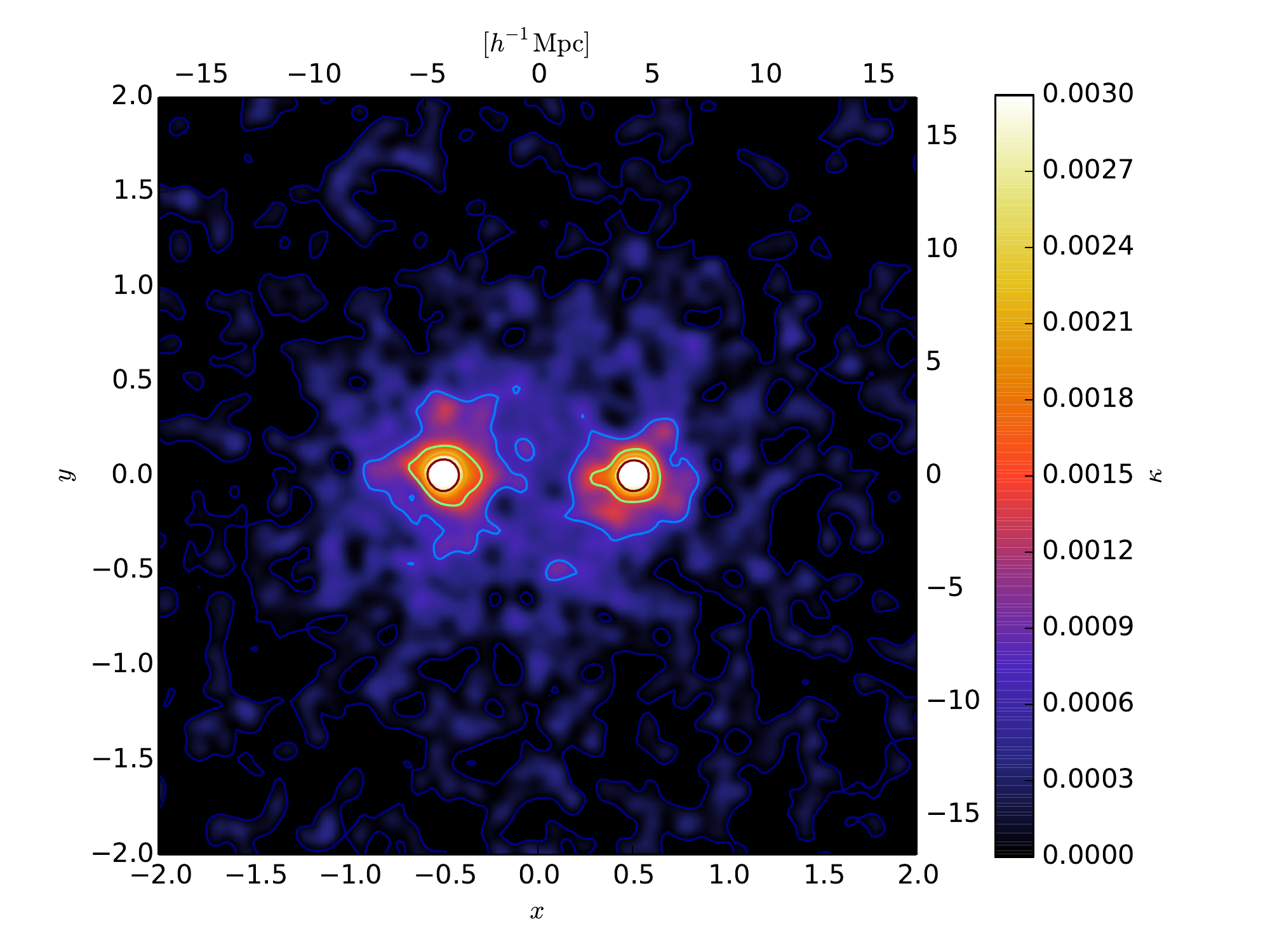}}
\caption{
The convergence ($\kappa$) map obtained from applying the Kaiser \& Squires inversion on to the shear map of Figure \ref{fig:shear_map}. A Gaussian smoothing filter of width 0.36 Mpc/h (0.04375 in units of $x,y$) has been applied to the convergence map for purposes of illustration.  (Top panel)  Reconstruction for physical LRG pairs. There is a clear sign of a mass bridge between the two LRGs. (Bottom panel) The same for the non-physical pairs of LRGs. The non-physical pairs lack the apparent filamentary feature between the LRGs.} 
\label{fig:convergence}
\end{figure*} 

The shear map after stacking pairs of LRGs with projected separations between 6$h^{-1}$Mpc and 10$h^{-1}$Mpc (average 8.23 $h^{-1}$Mpc) is shown in Figure \ref{fig:shear_map}. Figure \ref{fig:convergence} shows the resulting convergence map, with the upper panel showing the convergence around physical pairs of LRGs. The striking feature in this panel is the clear structure connecting the two physical LRGs. 
The lower panel of Figure \ref{fig:convergence} shows the convergence map from the lensing signal for non-physical LRG pairs,  in the same projected separation range ($6h^{-1}\mathrm{Mpc}\leq R_{\mathrm{sep}} < 10h^{-1}\mathrm{Mpc}$). A key feature of the lower panel is the lack of ``bridge'' between the two LRGs that is seen in the upper panel.

\begin{figure*}
  \centering
  \resizebox{2.0\columnwidth}{!}{\includegraphics{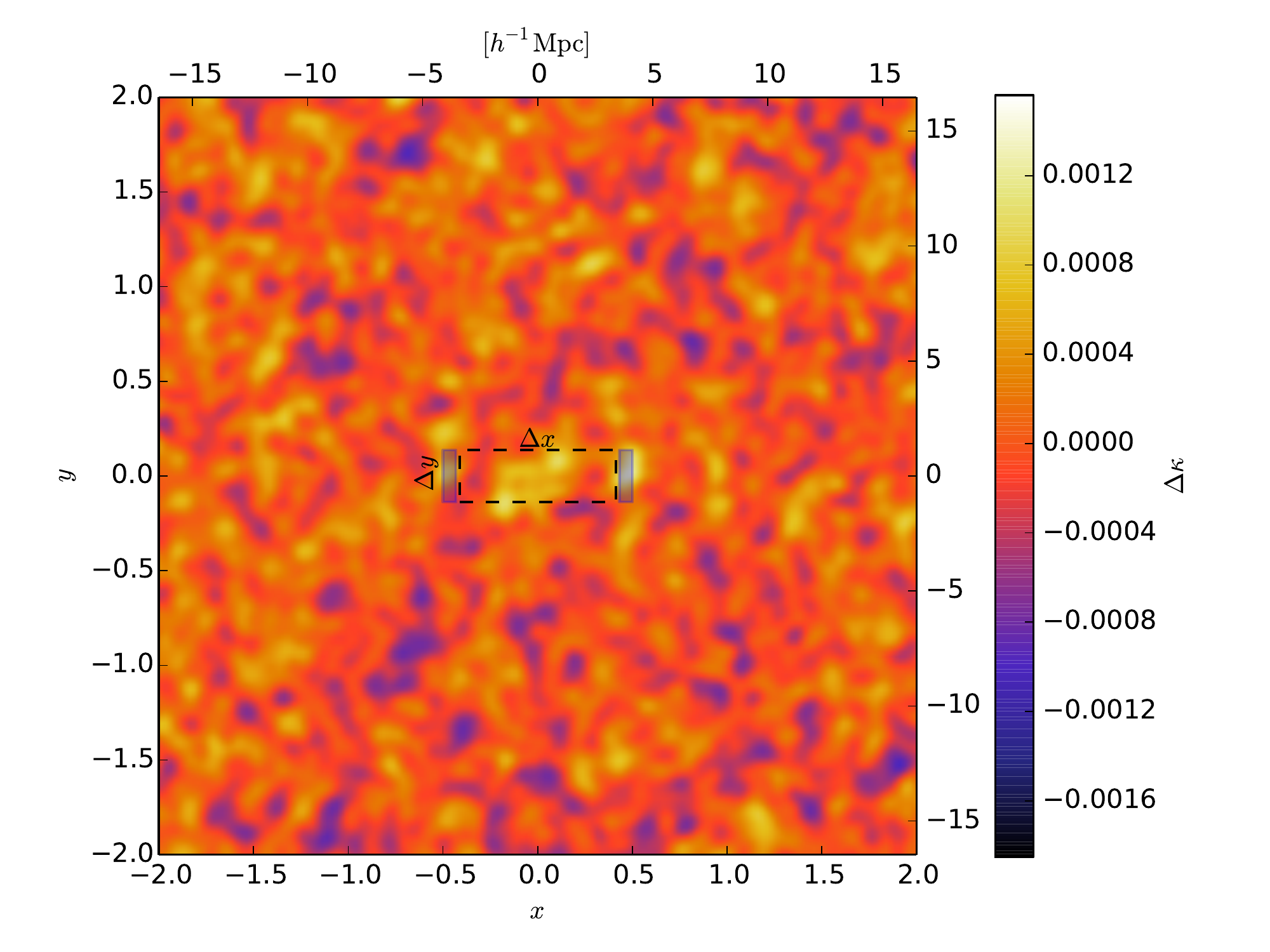}}
\caption{The result of subtracting the non-physical pair convergence map from that of the physical pairs. The shaded regions indicate where the regions within the LRG halos are excluded from the filament measurement. $\Delta x$ is fixed for all measurements, however $\Delta y$ varies, mapping out the convergence as a function of filament width, see Figure \ref{fig:kappa_v_dy}.} \label{fig:conv_sub}
\end{figure*}

To measure the residual filament signal, we begin by subtracting the convergence map of non-physical pairs (lower panel of Figure \ref{fig:convergence} from the convergence map of physical pairs (upper panel of Figure \ref{fig:convergence}). The result is shown in Figure \ref{fig:conv_sub}.  The excess surface mass density is clearly visible around the filament midpoint $(x,y) =(0,0)$.

To quantify the filament mass, we place a box of dimensions $\Delta x \times \Delta y$, representing the projected dimensions of the stacked filament (see Figure \ref{fig:conv_sub}), and measure the average excess convergence contained inside the box. After performing the direct subtraction of physical and projected pairs, there may be a small over- or under-subtraction of the convergence in the regions closest to the LRG positions due to a small differences in the mean physical and non-physical pair LRG masses. Moreover, the halos are likely to be elliptical and pointed along the line connecting the LRGs. Therefore, we wish to exclude from our definition of the filament, regions where the elliptical component of the LRG halos dominates the convergence. We note that some studies suggest that the $r_{200}$ of a dark matter halo may not be the optimal definition of its boundary with accreting matter extending well beyond $r_{200}$ \citep[e.g.\ ][]{OmaHudBeh13, 2015ApJ...810...36M}. To avoid including these LRG halo regions in the filament mass estimate, we only consider the filament to include points farther than $2r_{200}$ from either LRG.  The final width $\Delta x$ corresponds to 7.1 $h^{-1}$Mpc.

We estimate uncertainties via Monte Carlo simulations of the shape noise. Specifically,  we generate 1000 realizations by adding artificial scatter to the galaxy ellipticities consistent with shape noise. These noisy realizations are propagated to the $\kappa$ maps generated by the  \cite{1993ApJ...404..441K} method, and through the subtraction of non-physical pairs (the map of which has independent noise). Finally these uncertainties are propagated to the enclosed masses and mean $\kappa$ measurements discussed below.

\begin{figure}
  \centering
  \resizebox{0.95\columnwidth}{!}{\includegraphics{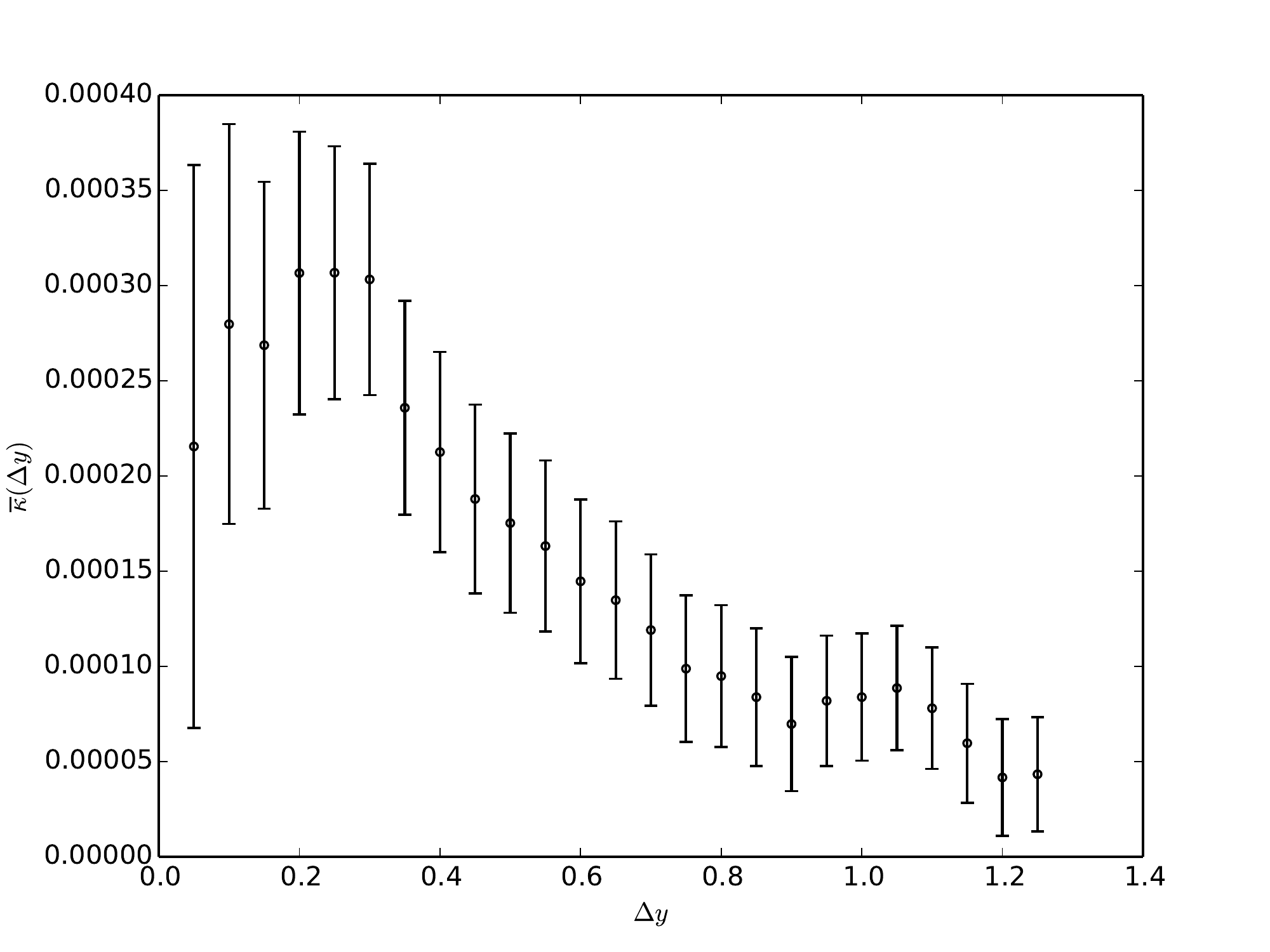}}
\caption{Mean convergence within a box of dimensions $\Delta x \times \Delta y$ as a function of increasing filament width, $\Delta y$. Note that because each point includes the convergence from smaller $\Delta y$, the plotted measurements are strongly correlated.} \label{fig:kappa_v_dy}
\end{figure} 

\begin{figure}
  \centering
  \resizebox{0.95\columnwidth}{!}{\includegraphics{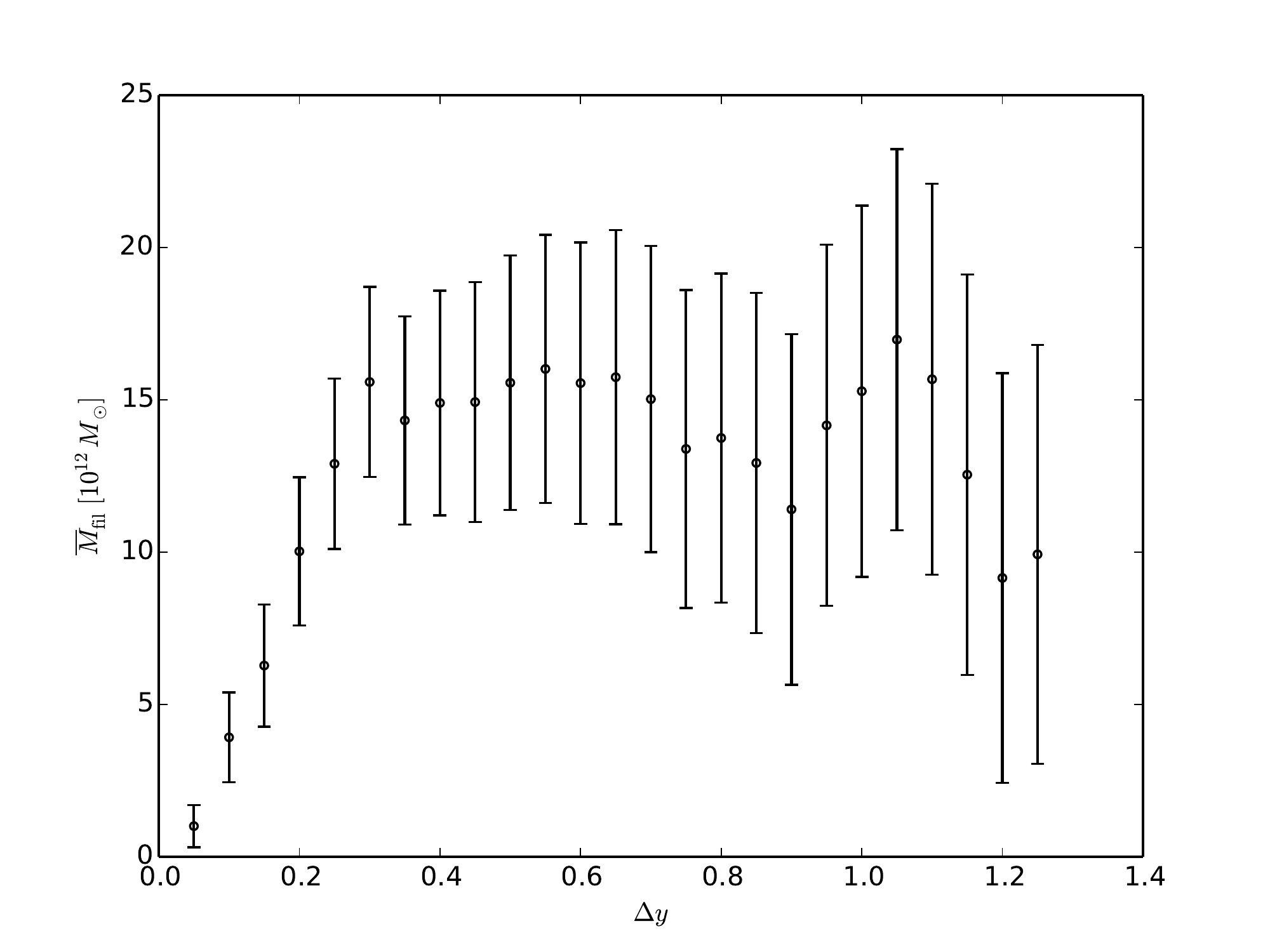}}
\caption[Average mass contained within the box defined by $\Delta x \times \Delta y$]{The average mass contained within the box defined by $\Delta x \times \Delta y$. Note that, as in \protect{Figure \ref{fig:kappa_v_dy}}, the measurements are strongly correlated. The filament mass shows no sign of increasing beyond $\Delta y = 0.3$ (corresponding to a physical width of $\sim 2.5h^{-1}\mathrm{Mpc}$). 
} \label{fig:av_mass}
\end{figure}

Figure \ref{fig:kappa_v_dy} shows the resulting mean convergence within the box as the width of the box, $\Delta y$, is increased. We then convert the convergence to a surface mass density using eq.\ (\ref{eqn:SigCrit_field}). It is then straightforward to calculated the average mass contained within the filament box, shown in Figure \ref{fig:av_mass}. From Figure \ref{fig:kappa_v_dy}, we see that the signal-to-noise peaks around $\Delta y = 0.3$ corresponding to a physical width of $2.5h^{-1}\mathrm{Mpc}$ at a significance of $\sim 5 \sigma$. The corresponding mass contained within the filament is $\overline{M}_{\mathrm{fil}} = (1.6 \pm 0.3)\times 10^{13}M_{\odot}$. The filament mass shows no sign of increasing beyond this $\Delta y$ so we adopt $2.5h^{-1}\mathrm{Mpc}$ as the fiducial width. 

The filament region has a projected length of $\sim 7 h^{-1}$ Mpc on the sky. We estimate that this corresponds to a true length $\sim 8 h^{-1}$ Mpc when the line of sight depth is included.  Assuming that the filament is a uniform density cylinder of length 8 $h^{-1}$Mpc and diameter 2.5$h^{-1}$Mpc, the corresponding excess density within the cylinder is then $\bar{\delta} = (\bar{\rho} - \rho_{b}) / \rho_{b} \sim 4$ where $\bar{\rho}$ is the mean density within the cylinder and $\rho_{b}$ is the background matter density.

The filament mass found here is a factor of a few less massive than the one reported by \cite{2012Natur.487..202D}, and about one order of magnitude less massive than the one reported by \cite{2015arXiv1503.06373H}. The difference in mass is likely due to the typical halo masses that connect the filament. The average host halo mass here is the order of $\sim 10^{13}M_{\odot}$, corresponding to a rich group rather than a massive cluster. In contrast, the host halos considered in \cite{2012Natur.487..202D} and \cite{2015arXiv1503.06373H} have masses of a few $10^{14}M_{\odot}$ up to $\sim 10^{15}M_{\odot}$ for \cite{2015arXiv1503.06373H}, corresponding to rich clusters of galaxies. 

The study of \cite{ClaMiyJai16} is similar to this work in the sense that it studies stacked filaments between LRGs. Their sample of LRGs was selected from SDSS-II, similar to the LRGs used in this study. Their paper does not provide a filament mass, perhaps because of the way in which the nulled filament is measured in their work makes it difficult to constrain directly.  They do analyze a set of stacked N-body filaments, which provide a reasonable fit to their signal. Examination of the convergence map of these filaments in their Figure 5, and allowing for the difference in $\Sigma_{\rm crit}$ suggests that the signals are comparable in mass.

\section{Modelling with the 3-Point Correlation Function}\label{sec:3PCF}

Figure \ref{fig:conv_sub} shows the stacked excess surface mass density around many pairs of LRGs.  It therefore does not correspond to an individual filament but an ensemble average of stacked filaments. To model it, we therefore consider the galaxy-galaxy-convergence (gg$\kappa$) 3-point correlation function (3PCF) derived from perturbation theory and developed in \cite{2014arXiv1402.3302C}. Here we summarize the key equations from that paper, to which the reader is referred for further details.

We are interested in the projected 3PCF around two dark matter halos at fixed locations $\vec{x}_{1}$ and $\vec{x}_{2}$, relative some matter at $\vec{x}_3$ which is denoted by%
	\begin{equation}\label{eqn:3pcf_1}
		\zeta_{\mathrm{gg}\kappa}(\vec{x}_{1}, \vec{x}_{2}, \vec{x}_{3})
		 = \langle \delta_\mathrm{g}(\vec{x}_{1})\delta_\mathrm{g}(\vec{x}_{2})\kappa(\vec{x}_3) \rangle,
	\end{equation}
where $\delta_\mathrm{g}$ is just the projected 3-dimensional galaxy overdensity.

Following \cite{2014arXiv1402.3302C}, the 3PCF can be derived from the bispectrum
given from perturbation theory by \cite{2002PhR...367....1B}:
	\begin{eqnarray}\label{eqn:bispec_bern}
		B(\vec{k}_1, \vec{k}_2, \vec{k}_3) &=& \left [ \frac{10}{7} + \left ( \frac{k_1}{k_2} + \frac{k_2}{k_1} \right ) \right.
		\left. \frac{\vec{k}_1\cdot \vec{k}_2}{k_1 k_2} + \right. \nonumber \\
		&&\left. \frac{4}{7}\frac{(\vec{k}_1\cdot \vec{k}_2)^2}{k^2_1 k^2_2}  \right ]
		P^{\mathrm{L}}_{\mathrm{m}}(k_1)P^{\mathrm{L}}_{\mathrm{m}}(k_2) \nonumber \\
		&+& \mathrm{permutations} ,
	\end{eqnarray}
where $P^{\mathrm{L}}_{\mathrm{m}}(k)$ is the linear matter power spectrum.
	\begin{eqnarray}\label{eqn:3pcf_6}
		\zeta_{\mathrm{gg}\kappa}(\vec{x}_{1}, \vec{x}_{2}, \vec{x}_{3})
		&=& \frac{\Sigma^{-1}_{\mathrm{crit}}(\chi_{\mathrm{L}}, \chi_{\mathrm{s}})}
		{\sqrt{2\pi}\sigma_{\mathrm{LRG}}} \rho_{\mathrm{crit},0}\Omega_{\mathrm{m},0}
		b^2\frac{1}{(2\pi)^3} \nonumber \\
		&\times& \int^{\infty}_0dk_1\int^{\infty}_0dk_2\int^{2\pi}_0d\phi k_1k_2 B(k_1, k_2, -k_{12}) \nonumber \\
		&\times& \mathrm{J}_0(\sqrt{\alpha^2 + \beta^2}),
	\end{eqnarray}
where $b$ is the linear bias of LRGs and $\sigma_{\mathrm{LRG}}$ is the typical separation of LRGs along the line of sight, converted to physical units.
This integral can be evaluated numerically for a given separation bin as described in \S\ref{sec:sdss}.

\begin{figure*}
  \centering
  \resizebox{2.0\columnwidth}{!}{\includegraphics{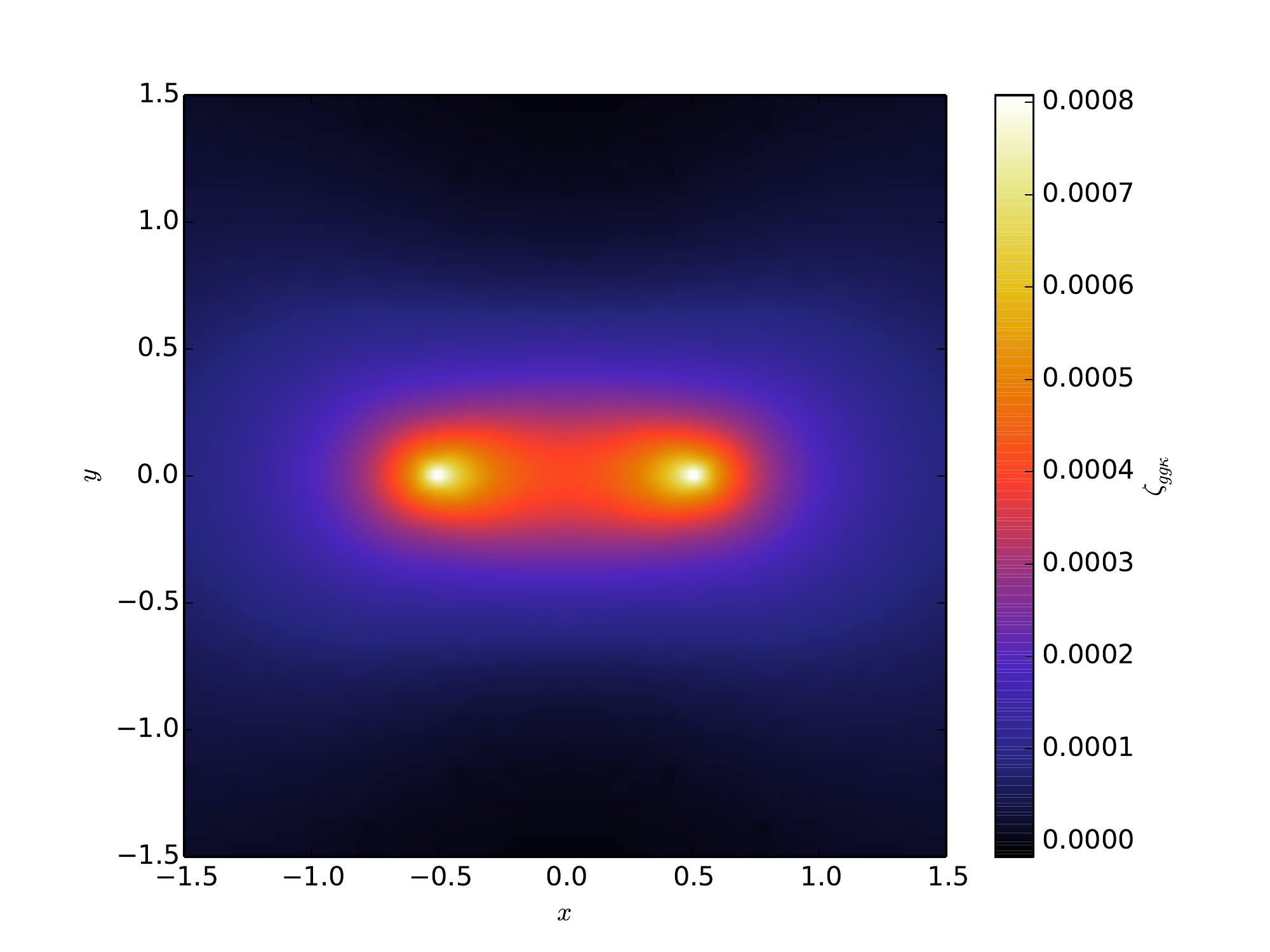}}
\caption[The model 3-point galaxy-galaxy-$\kappa$ correlation function]{The three-point galaxy-galaxy-convergence calculated by numerically integrating equation (\ref{eqn:3pcf_6}) for the separation range $6h^{-1}\mathrm{Mpc} < R < 10h^{-1}\mathrm{Mpc}$. This is plotted in the standardized coordinate system defined in \S\ref{sec:coords}.} \label{fig:3pcf_conv}
\end{figure*}

The three-point convergence map generated for projected separations $6h^{-1}\mathrm{Mpc}\leq R_{\mathrm{sep}} < 10h^{-1}\mathrm{Mpc}$ is shown in Figure \ref{fig:3pcf_conv}. Here we have used a linear bias, $b$, of 2 \citep{2015ApJ...806....2M, 2014MNRAS.440.2222T} and we follow \cite{2014arXiv1402.3302C} to estimate the r.m.s.\ line of sight separation of LRGs $\sigma_{\mathrm{LRG}} = 8 h^{-1}$ Mpc. It is important to take care to ensure that the resulting convergence map is in physical units; the integral in eq.\ (\ref{eqn:3pcf_6}) is done over comoving coordinates, which introduces an additional factor of $(1+z_{l})^{-2}$. The factor of $\Sigma_{\mathrm{crit}}$ was measured from the data according to eq.\ (\ref{eqn:SigCrit_field}).
\subsection{Results}
\begin{figure*}
  \centering
  \resizebox{1.8\columnwidth}{!}{\includegraphics{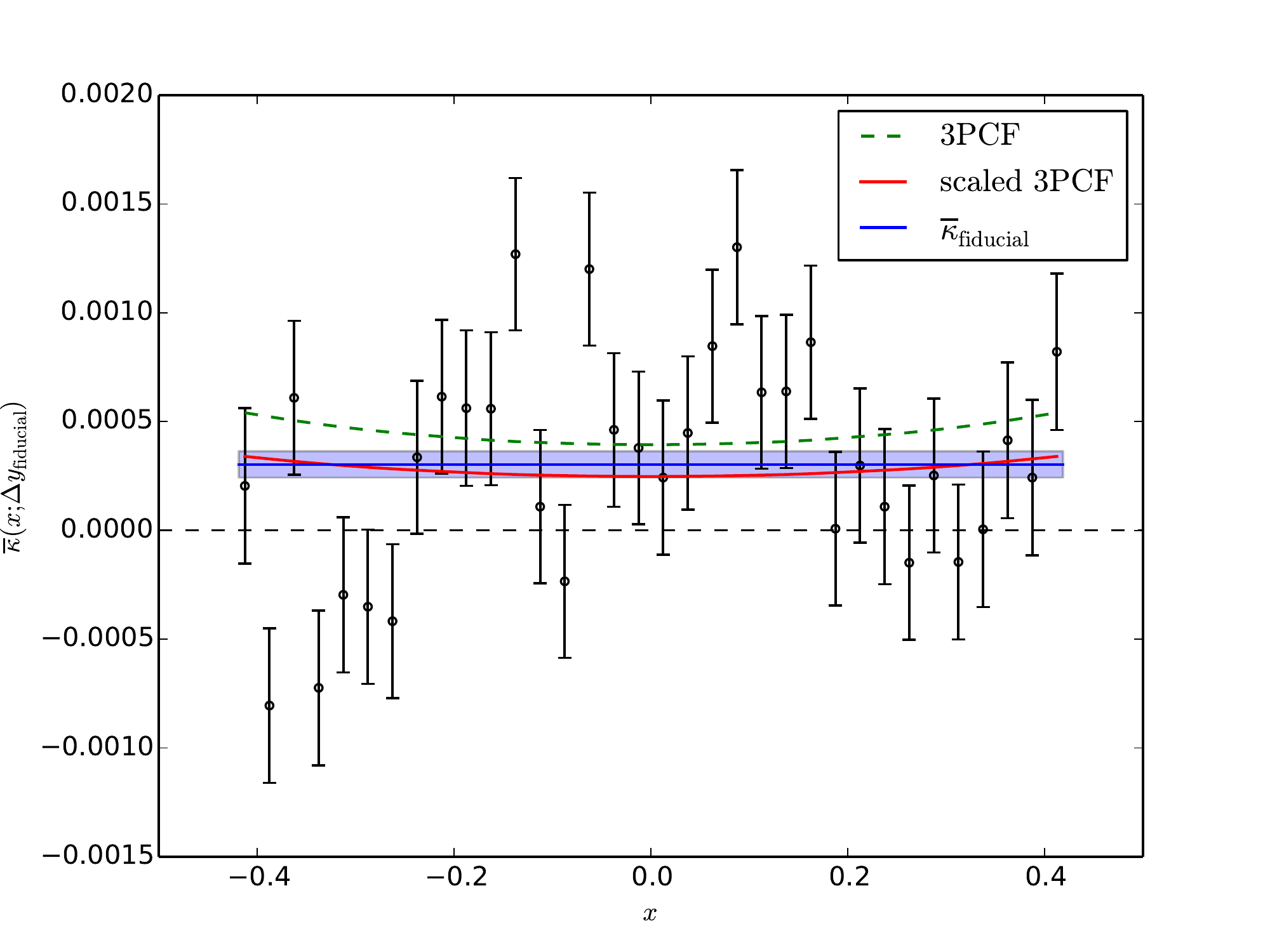}}
\caption{The resulting convergence profile along the $x$-axis for the method of non-physical pair subtraction. This is done for the fiducial $\Delta y = 2.5 h^{-1}\mathrm{Mpc}$ calculated in \S\ref{sec:emp_results}, with the average convergence for that particular box plotted in blue. The model three-point correlation function is plotted in green and the 3PCF model scaled down by a factor of $0.6$ is plotted in red. For comparison the mean convergence within the fiducial width is also shown in blue.} \label{fig:3pcf_notscaled}
\end{figure*}

As discussed in \S\ref{sec:emp_results}, the filament signal showed no significant increase beyond the fiducial width of $\Delta y = 0.3 \sim 2.5 h^{-1}$ Mpc, so we adopt this width to compare the filament data with the 3PCF model. Figure \ref{fig:3pcf_notscaled} shows convergence data binned along $x$-axis as well as the the three-point correlation function, averaged over the fiducial width. Also shown is the total averaged convergence within the filament box. At a glance, it appears the three-point function fits the data well, however the model lies slightly above the best fitting value.  While the model appears to be a good fit to the central filament region ($x \sim 0$), the data do not appear to show the excess around the two LRGS ($x = \pm 0.5$) that is both predicted by the 3PCF and seen in simulations \citep{2005MNRAS.359..272C}. Neglecting this and simply performing a least squares fit to the entire $x$ range suggests that the model  overestimates the data by a factor of $\sim 1.6$.
\subsection{Discussion}
The required re-scaling of the 3PCF model is relatively small, being on the order of uncertainty in the data (roughly $20$\%). 
It is possible that the model is an overestimate, due to an \emph{under}estimate of the effect of LRG peculiar velocities. 
In calculating the three-point correlation function model, we followed \cite{2014arXiv1402.3302C}, and parameterized the line-of-sight separation of the two LRGs by a Gaussian distribution with width  $\sigma_{\mathrm{LRG}} = 8 h^{-1}$ Mpc. This separation in redshift space includes both the peculiar velocities of each LRG in the pair and the Hubble flow. The peculiar velocities are difficult to model since they include contributions from relative infall motions as well as ``thermal'' motions of the LRGs themselves within their host halos. This model could be improved by using a more physically motivated distribution, using two-point statistics, as well as careful calibration from N-body studies. 
\section{Conclusions}\label{sec:conc}
The formation of a filamentary structure that connects high density collapsed regions of the universe is a prediction from simulations that has only recently become detectable observationally. In this work, we have detected a stacked filamentary structure between SDSS-III/BOSS LRGs using the CFHTLenS data set. The filament detection is significant at the $5\sigma$ level, with a mass of $(1.6 \pm 0.3) \times 10^{13} M_{\odot}$ for a box of fiducial physical dimensions, $7.1\times 2.5 h^{-1}\mathrm{Mpc}$.  The three-point correlation function was used as a model for the stacked filament, derived from the perturbation theory bispectrum. We have shown that the predictions of the three-point correlation function are in reasonable agreement to the data.

The goal of this study was to detect filaments using weak lensing, but also to serve as a foundation for future filament studies. We have developed a simple method of stacking filaments that can be applied to any weak lensing dataset, provided one has obtained redshifts for groups and clusters of galaxies through spectroscopy.  Upcoming surveys such as the DES \citep{2005astro.ph.10346T} will obtain ellipticities over 5000 square degrees to approximately the same depth as CFHTLenS. Presently there is little spectroscopy in the DES footprint, however. Other surveys such as SuMIRe/Hyper-Suprime Cam\footnote{http://sumire.ipmu.jp/en/}, 2dFLenS \citep{BlaAmoChi16} and the Canada-France Imaging Survey\footnote{www.cfht.hawaii.edu/Science/CFIS} will greatly increase the overlap between spectroscopic foreground lens samples and deep samples of background source galaxies.  As well as new ground-based surveys, planned space-based missions, such as Euclid \citep{LauAmiArd11} or WFIRST \citep{2015arXiv150303757S} have the potential to measure the ellipticities and photometry of billions of galaxies . With increases in statistical power it will be come possible to study the nature of filaments as a function of other properties such as halo mass, separation and redshift.

\section*{Acknowledgments}

We acknowledge useful discussions with Joseph Clampitt. We also acknowledge the substantial efforts of both the CFHT staff in implementing the Legacy Survey, and of the CFHTLenS team in preparing catalogues of galaxy ellipticities and photometric redshifts. MJH acknowledges support from NSERC.

Based on observations obtained with MegaPrime/MegaCam, a joint project of CFHT and CEA/IRFU, at the Canada-France-Hawaii Telescope (CFHT) which is operated by the National Research Council (NRC) of Canada, the Institut National des Science de l'Univers of the Centre National de la Recherche Scientifique (CNRS) of France, and the University of Hawaii. This work is based in part on data products produced at Terapix available at the Canadian Astronomy Data Centre as part of the Canada-France-Hawaii Telescope Legacy Survey, a collaborative project of NRC and CNRS.

\end{document}